\documentclass[final,authoryear,5p,times,twocolumn]{elsarticle}

\makeatletter
\def\blfootnote{\xdef\@thefnmark{}\@footnotetext}
\makeatother

 \usepackage{graphicx}
\usepackage{amssymb}
\usepackage{bm}
\usepackage{amsmath}
\usepackage{fancyvrb}
\usepackage{color} % Delme
\usepackage{breqn}
\usepackage{rotating}
\usepackage{upgreek}
\usepackage{graphbox}
\usepackage{hyperref}

\usepackage{color} % Delme

\newcommand{\simnot}{\mathord{\sim}}

\journal{Astronomy and Computing}

\begin{document}

\begin{frontmatter}

%% Title, authors and addresses

%% use the tnoteref command within \title for footnotes;
%% use the tnotetext command for the associated footnote;
%% use the fnref command within \author or \address for footnotes;
%% use the fntext command for the associated footnote;
%% use the corref command within \author for corresponding author footnotes;
%% use the cortext command for the associated footnote;
%% use the ead command for the email address,
%% and the form \ead[url] for the home page:
%%
\title{Cleaning radio interferometric images using a spherical wavelet decomposition} %%\tnoteref{label1}}
%% \tnotetext[label1]{}
\author{Chris~J.~Skipper}
\ead{chris.skipper@manchester.ac.uk}
\author{Anna.~M.~M.~Scaife} %%\corref{cor1}\fnref{label2}}
%% \ead[url]{home page}
%% \fntext[label2]{}
%% \cortext[cor1]{}
\address{Jodrell Bank Centre for Astrophysics, Alan Turin Building, The University of Manchester, Manchester, M13 9PL, UK} %
\author{Jason~D.~McEwen}
\address{Mullard Space Science Laboratory (MSSL), University College London, Surrey, RH5 6NT}%\fnref{label3}}
%% \fntext[label3]{}

%% use optional labels to link authors explicitly to addresses:
%% \author[label1,label2]{<author name>}
%% \address[label1]{<address>}
%% \address[label2]{<address>}

\begin{abstract}
The deconvolution, or cleaning, of radio interferometric images often involves computing model visibilities from a list of clean components, in order that the contribution from the model can be subtracted from the observed visibilities. This step is normally performed using a forward fast Fourier transform (FFT), followed by a `degridding' step that interpolates over the uv plane to construct the model visibilities. An alternative approach is to calculate the model visibilities directly by summing over all the members of the clean component list, which is a more accurate method that can also be much slower. However, if the clean components are used to construct a model image on the surface of the celestial sphere then the model visibilities can be generated directly from the wavelet coefficients, and the sparsity of the model means that most of these coefficients are zero, and can be ignored. We have constructed a prototype imager that uses a spherical-wavelet representation of the model image to generate model visibilities during each major cycle, and find empirically that the execution time scales with the wavelet resolution level, $J$, as ${\mathcal{O}(1.07^{J})}$, and with the number of distinct clean components, ${N_{\rm C}}$, as ${\mathcal{O}(N_{\rm C})}$. The prototype organises the wavelet coefficients into a tree structure, and does not store or process the zero wavelet coefficients.
\end{abstract}

\begin{keyword}
%% keywords here, in the form: keyword \sep keyword
techniques: interferometric

%% MSC codes here, in the form: \MSC code \sep code
%% or \MSC[2008] code \sep code (2000 is the default)

\end{keyword}

\end{frontmatter}

% \linenumbers

%% main text
\section{Introduction}

\blfootnote{\textcopyright~2019 This manuscript version is made available under the \mbox{CC-BY-NC-ND 4.0} license \url{http://creativecommons.org/licenses/by-nc-nd/4.0/}} The advent of the radio interferometer in astronomy opened a new high-resolution window on the universe, and the scale and ambition of astronomical interferometers continues to grow exponentially. The next generation of interferometers, such as the forthcoming Square Kilometre Array (SKA; \citealt{Schilizzi2008, Dewdney2009}), which is due to begin construction soon, and its precursors, MeerKAT \citep{Jonas2009, Booth2012} and ASKAP \citep{Johnston2008}, which are currently being brought online, will require new innovations, algorithms and hardware capabilities in order to meet the huge demand for compute capability that they will present.

A radio interferometer measures the interference pattern produced by two or more apertures, and therefore samples the sky in the Fourier domain. To recover images from these visibilities is an entirely computational problem, and one which scales (depending upon which step is being performed) in proportion to the number of baselines between pairs of antennas, the size of the image required, or the length of the observation. The greatest computational expense when constructing images from radio data is arguably the cleaning, or deconvolution, stage. Since, in interferometry, coverage of the Fourier plane is always incomplete it follows that the synthesised beam of an interferometer is imperfect compared to that of a filled-aperture detector. Correcting for these imperfections is a slow process.

\begin{figure*}
        \centering
	\includegraphics[clip, trim=0cm 3.8cm 0cm 1.2cm, width=140mm]{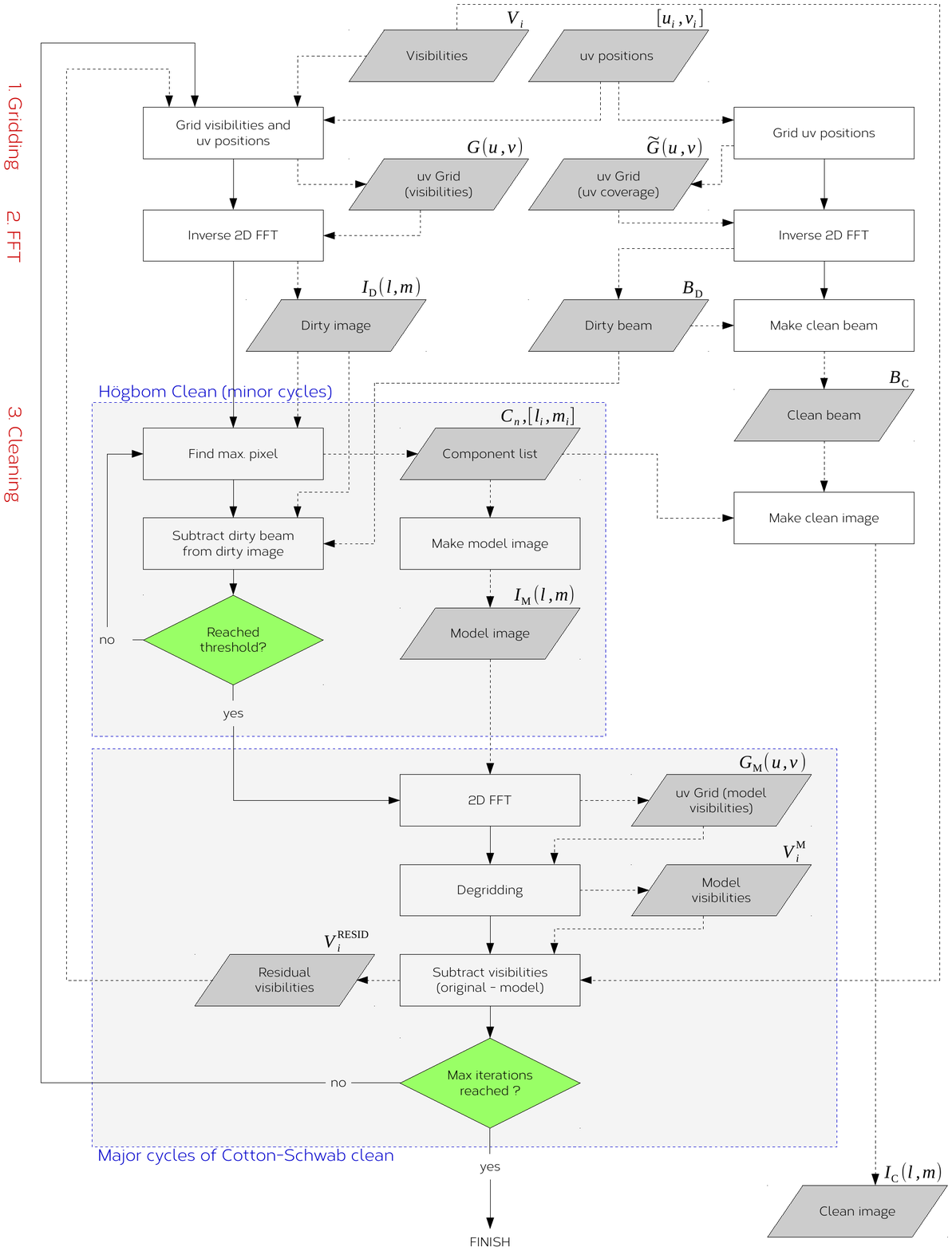}
        \caption{A typically radio-imaging pipeline, demonstrating the gridding, FFT and cleaning (deconvolution) stages. The minor cycle stage performs a H\"ogbom clean, in which the dirty image is cleaned down to some threshold and a model image is built up from the identified components. In the major-cycle stage, a two-dimensional FFT is applied to the model image to transform to the uv domain, and an interpolation algorithm calculates each model visibility. The model image and clean image can be built up gradually with each iteration of H\"ogbom clean.}
        \label{fig-flow-chart-cotton-schwab}
\end{figure*}

To `clean' radio images we would normally invoke an iterative process of subtracting instances of the synthesised beam from our uncleaned, or dirty, image, and instead build up a clean component list or model image \citep{Hogbom1974}. The flowchart in Fig.~\ref{fig-flow-chart-cotton-schwab} shows the construction of a component list from iterations of H\"ogbom clean. Following this process of iterations we would then recalculate our visibilities based upon the model image, subtract these model visibilities from the observed visibilities, and generate a new dirty image which will be cleaned with a new set of deconvolution iterations.

Model visibilities can be calculated directly from the component list by summing the contribution from each component separately, or more commonly from the model image by using a forward fast Fourier transform (FFT), followed by an interpolation process known as degridding. The former scales in performance with the number of visibilities and the number of clean components, while the latter scales with the size of the image in the FFT stage, and the number of visibilities in the degridding stage. \cite{McEwen2008} developed a fast method to simulate visibilities which uses a wavelet representation over the surface of the celestial sphere. Their technique takes advantage of the sparse representation of the sky in the wavelet basis, and recognises that many of the coefficients are zero or nearly zero. Their algorithm uses hard thresholding to determine which coefficients to retain and which to reject, set by defining a fixed proportion of coefficients to retain, and alternatively, by using an annealing strategy. The authors report that visibilities could be computed accurately when fewer than one per cent of the coefficients were retained.

When generating model visibilities from the model image using FFT and degridding steps we are using a tangent-plane approximation, which is only valid for small fields of view (FOV). As the FOV becomes larger, and the area of the sky within the image region encompasses more of the curvature of the celestial sphere, this approximation becomes less reliable. In contrast, a spherical representation of the model image, using wavelets or spherical harmonics, can be constructed over the whole of the sphere, and the visibilities can be generated with no reliance on a tangent-plane approximation, making a spherical representation ideal for reconstructing model visibilities for use in wide-field imaging deconvolution. The technique may therefore be of particular use to instruments designed for deep surveying of the sky, to which large FOVs are an essential requirement. \cite{McEwen2008} demonstrated that using a fast wavelet method is much more efficient at generating visibilities than using a spherical harmonic representation, so here we confine our discussion purely to the spherical wavelet method.

In this paper we have implemented the theory described by \cite{McEwen2008}, and constructed a full imaging propotype which uses a spherical-wavelet representation to compute visibilities during the major cycles of the cleaning stage. We describe in detail the way in which cleaning is normally performed in radio astronomy, and how spherical wavelets fit in to this process.

\section{Radio imaging in astronomy}

\subsection{Overview}

\begin{figure}
        \centering
	\includegraphics[clip, width=65mm]{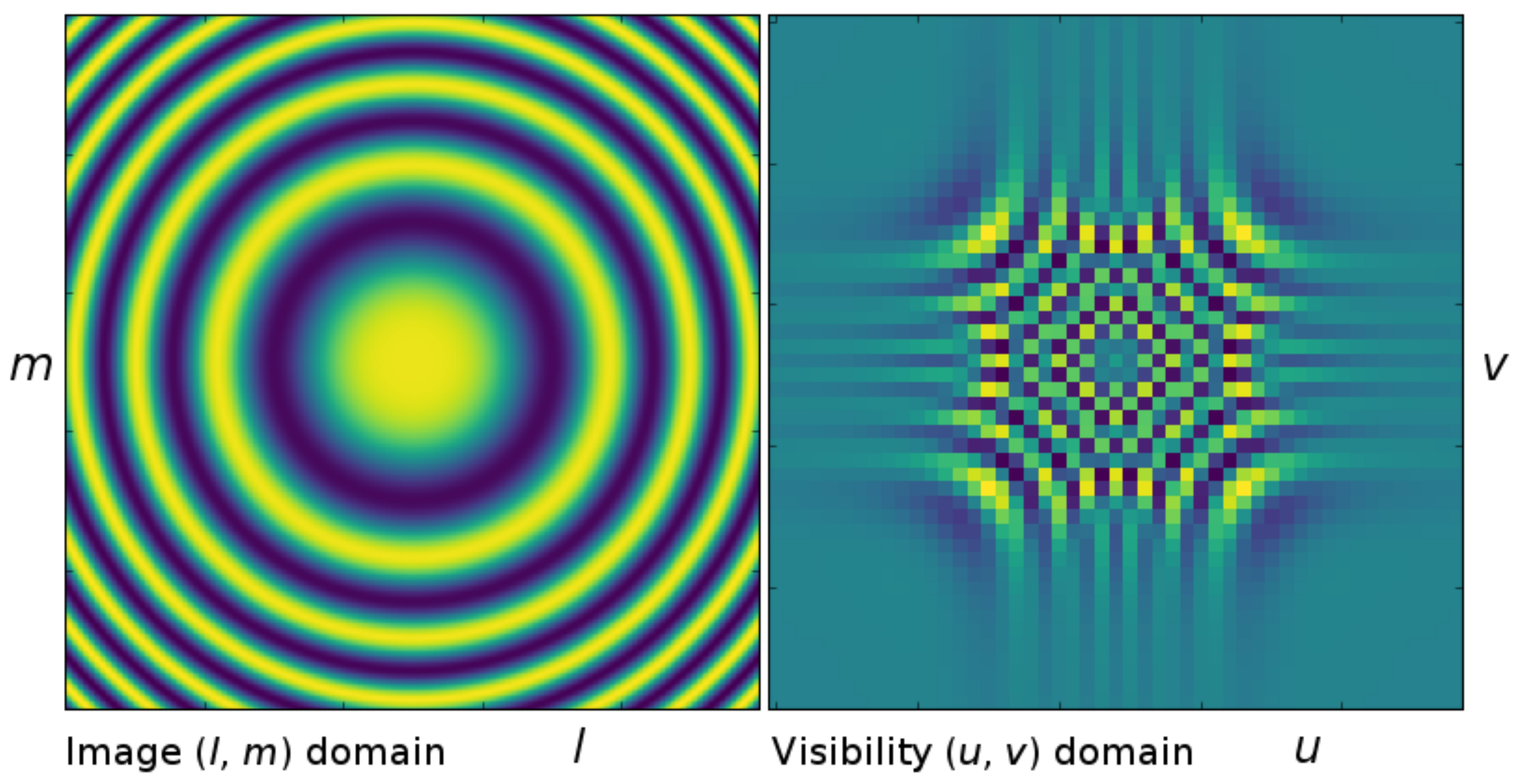}
        \caption{Example w-kernel for degridding, ${K^{j}_{u,v}}$, shown in the image domain (left panel) and uv domain (right panel).}
        \label{fig-w-kernel}
\end{figure}

High-resolution radio imaging is generally obtained through the use of radio interferometers, which are arrays of radio antennas arranged in some configuration such that each pair discretely samples a unique Fourier component of the radio sky. Together, the correlated amplitude from each pair of antennae (the 'visibilities', ${V_i}$), and the $u$ and $v$ components of their baseline vector, form the two-dimensional Fourier transform of the sky image, ${I( l, m )}$, where $l$ and $m$ are directional cosines on the sky. The full baseline vector (${\bm{B}_i = [u_i, v_i, w_i]}$, where ${u_i, v_i, w_i \in \mathbb{R}}$), includes the $w$ term, and is orientated such that the $w$-axis points towards the phase position on the sky, and the $u$ and $v$ axes form the uv plane. The visibilities are related to the sky image by
\begin{equation}
	\begin{split}
	V_{j} = \int \int \frac{I( l, m )}{\sqrt{1 - l^2 - m^2}} \, {\rm exp}\bigg[ \, i2\pi \left( u_j l + v_j m + \right. \\
	\left. w_j \left( \sqrt{1 - l^2 - m^2} - 1 \right) \right) \, \bigg] \, {\rm d}l \, {\rm d}m,
	\end{split}
	\label{eqn-visibility}
\end{equation}
and for small fields of view, where ${1 - l^2 -m^2 \approx 1}$, this equation reduces to
\begin{equation}
	V_{j} = \int \int {I( l, m ) \, {\rm exp}\bigg[ \, i2\pi \left( u_j l + v_j m \right) \, \bigg] \, {\rm d}l \, {\rm d}m},
	\label{eqn-visibility-small-fov}
\end{equation}
i.e. a two-dimension Fourier transform of the sky image ${I( l, m )}$, while for larger fields of view the components of the baseline vectors perpendicular to the uv plane (the $w$ terms) require (see \citealt{Cornwell2012,Pratley2018}) that visibilities recovered through a Fourier transform are convolved with a 'w-kernel', ${K^{j}_{u,v}}$, given by (and shown graphically in Fig. \ref{fig-w-kernel})
\begin{equation}
	K^{j}_{u,v} = \mathcal{F} \Bigg\{ \frac{1}{\sqrt{1 - l^2 - m^2}} \, {\rm exp}\bigg[ \, i2\pi w_j \left( \sqrt{1 - l^2 - m^2} - 1\right) \, \bigg] \Bigg\}.
\end{equation}

As the visibility equation (Eqn.~\ref{eqn-visibility}) cannot be used to analytically determine the sky image, ${I( l, m )}$, from a set of known visibilities it follows that fast Fourier transforms are almost always employed in the recovery of ${I( l, m )}$. The sky image is a convolution of the components present within the field of view and the synthesised beam pattern. In practise, the uv plane is only partially sampled, and the synthesised beam pattern (the 'dirty' beam, ${B_{\rm D}}$) is imperfect. The process of recovering the clean sky image ${I_{\rm C}( l, m )}$ traditionally follows a number of well-established steps:

\begin{enumerate}
	\item Gridding: each visibility ${V_i}$ is placed on a discrete grid, ${G( u, v )}$, at its associated uv position ${[u_i, v_i]}$. The visibilities are usually convolved with a 2D weighting function, such as a Gaussian or prolate spheroidal function, and the w-kernel function (if required). To generate the dirty beam, the visibilities are gridded again at the same positions, but this time their values are all set to unity; the resulting grid, ${\hat{G}( u, v )}$, is the uv coverage.
	\item FFT: the 'dirty' image of the sky, ${I_{\rm D}( l, m )}$, is recovered using a two-dimensional inverse FFT of the gridded visibilities, i.e. ${\mathcal{F}^{-1}\{ G( u, v ) \}}$, and the dirty beam, ${B_{\rm D}}$, is similarly recovered using ${\mathcal{F}^{-1}\{ \hat{G}( u, v ) \}}$.
	\item Cleaning/deconvolution: A cleaning algorithm is applied, that removes the convolution of the dirty beam by building a list of components directly from the dirty image. A clean beam, ${B_{\rm C}}$, is constructed by fitting the dirty beam with a two-dimensional Gaussian function, rotated by some angle. The clean image, ${I_{\rm C}( l, m )}$, is then recovered by the convolution of the component list with the clean beam.
\end{enumerate}

This process is represented as a flow chart in Fig.~\ref{fig-flow-chart-cotton-schwab}.

\subsection{The cleaning algorithm}
\label{sec-radio-imaging-cleaning}

The basis of deconvolving the dirty beam from the sky image is an iterative algorithm called the CLEAN algorithm \citep[or simply H\"ogbom clean;][]{Hogbom1974}, and involves repeatedly identifying the pixel with the maximum value, ${C_i}$, at position ${[l_i, m_i]}$ (where ${l_i, m_i \in \mathbb{Z}}$) in the dirty image, and subtracting the dirty beam from that position, such that, for iteration $n$,
\begin{equation}
	I_{\rm D}^{(n + 1)}( l, m ) = I_{\rm D}^{(n)}( l, m ) - g C_n \big[ B_{\rm D} \circledast \delta( l_n, m_n ) \big],
\end{equation}
where $g$ is the gain parameter, typically fixed to $0.1$ or $0.2$, and $\delta$ is the Dirac delta function; with each iteration the algorithm builds up a list of clean components and positions, $C_n$ and ${[l_n, m_n]}$, a model image
\begin{equation}
	I_{\rm M}( l, m ) = \sum_i g C_i \, \delta( l_i, m_i ),
\end{equation}
and a clean image
\begin{equation}
	I_{\rm C}( l, m ) = \sum_i g C_i \big[ B_{\rm C} \circledast \delta( l_i, m_i ) \big].
\end{equation}

More advanced cleaning algorithms are built upon the H\"ogbom clean, such as Clark clean \citep{Clark1980}, and Cotton-Schwab clean \citep{Schwab1984}. The latter introduces minor and major cycles, with the minor cycles provided by the iterations of the H\"ogbom clean, and the major cycles, which are performed every time the minor-cycle stage reaches some cut-off threshold, described as follows:

\begin{enumerate}
	\item FFT and `degridding': A forward FFT is performed on the model image, ${I_{\rm M}( l, m )}$, to transform into the uv domain. An interpolation algorithm then calculates each `model' visibility, ${V_i^{\rm M}}$, at its known position ${[u_i, v_i]}$, in a process called degridding. The more accurate alternative is to calculate each visibility directly from Eqn.~\ref{eqn-visibility} (using the model image, ${I_{\rm M}( l, m )}$), but such an approach is only computationally efficient if the number of clean components is low.
	\item Subtract visibilities: the model visibilities are subtracted from the observed visibilities to give the residual visibilities, such that ${V_i^{\rm RESID} = V_i - V_i^{\rm M}}$.
	\item Gridding and FFT: the residual visibilities, ${V_i^{\rm RESID}}$, are placed on the uv grid, and an inverse FFT is performed to generate a new dirty image.
\end{enumerate}

\section{Deconvolution using spherical wavelets}

\subsection{Background on wavelets}
\label{sec-wavelets-background}

A wavelet series, such as the Haar wavelet \citep{Haar1910}, is a representation of a function, ${F(t)}$, in terms of an alternative basis or dictionary (hereafter we use the term basis for brevity). The basis is typically described in terms of a scaling function, $\phi$, and a series of wavelet functions, ${\psi_{j}}$, at scales $j$. These functions, together with their coefficients, can be used to reconstruct the original function ${F(t)}$. In the explanation that follows, we restrict our discussion to the discrete wavelet transform, in which our original function, ${F(t)}$ is sampled discretely at regular time intervals, which we denote ${F_k}$.

The coefficients of the scaling and wavelet functions (denoted ${\lambda_{j, k}}$ and ${\gamma_{j, k}}$ respectively, for resolution $j$ and index $k$) are calculated at multiple resolutions, such that at resolution $j$ (where ${j \ge 1}$) there are ${2^{j - 1}}$ scaling coefficients and ${2^{j - 1}}$ wavelet coefficients. At the highest resolution, ${j = J}$, the scaling coefficients of the Haar wavelet store the values of the discretely sampled function, i.e. ${\lambda_{J, k} = F_k}$, and at all resolutions $j < J$ the scaling coefficients are calculated using
\begin{equation}
	\lambda_{j, k} = \frac{1}{2} \left( \lambda_{j + 1, 2k} + \lambda_{j + 1, 2k + 1} \right).
\end{equation}

At the lowest resolution, ${j = 1}$, there is only one scaling coefficient, which stores the mean value of the function ${F_k}$. The wavelet coefficients are defined only for resolutions ${j = 1}$ to ${j = J - 1}$, and store the differences between the scaling coefficients of adjacent resolutions, i.e. ${\gamma_{j, k} = \lambda_{j + 1, 2k} - \lambda_{j, k} = -\left( \lambda_{j + 1, 2k + 1} - \lambda_{j, k} \right)}$. The wavelet coefficients are more usually written
\begin{equation}
	\gamma_{j, k} = \frac{1}{2} \left( \lambda_{j + 1, 2k} - \lambda_{j + 1, 2k + 1} \right).
\end{equation}

The complete wavelet transform is described by the scaling coefficient at resolution ${j = 1}$, i.e. ${\lambda_{1, 0}}$, and the wavelet coefficients at all levels, ${\gamma_{j, k}}$.

\subsection{Wavelets on the sphere}

The Haar wavelet transform can be applied to two-dimensional functions, such as images, as well as one-dimensional time series. Extensions of Haar wavelets to the surface of a sphere \citep{Tenorio1999, Barreiro2000, McEwen2008, McEwen2011} rely upon hierarchical pixelation schemes. Following \cite{McEwen2008} we base our pixelation upon the {\sc healpix} pixelation scheme \footnote{http://healpix.jpl.nasa.gov} \citep{Gorski2005}, but for the purposes of the research in this paper choose to develop our own software rather than implement any third-party code.

\begin{figure}
        \centering
	\includegraphics[clip, width=85mm]{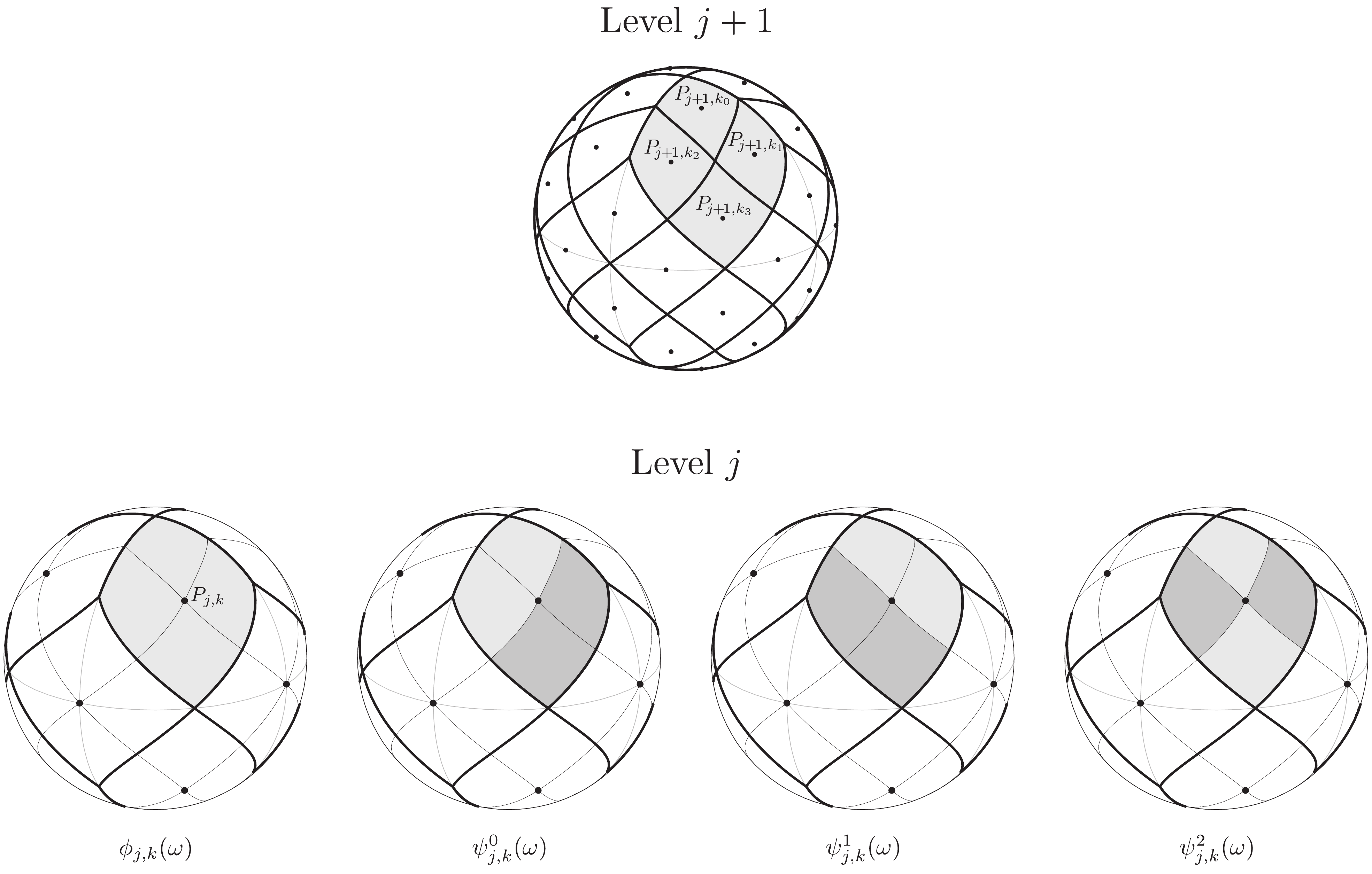}
        \caption{Haar scaling function ${\phi_{j, k}(\bm{\hat{s}})}$ and wavelets ${\psi^m_{j, k}(\bm{\hat{s}})}$ (image from \citealt{McEwen2008}). Negative, zero, and positive constant values are indicated with dark shaded regions, unshaded regions, and light shaded regions respectively, and the scaling function and wavelets at resolution $j$ and index $k$ are non-zero only on pixel ${P_{j, k}}$. Each pixel at resolution $j$ is divided into four pixels at level ${j + 1}$.}
        \label{fig-haar-wavelets}
\end{figure}

Like {\sc healpix}, our software divides the sphere into twelve equal-sized pixels at resolution ${j = 1}$, and at each higher resolution ${j + 1}$ we divide each pixel into four new, equal-sized pixels. This division into four new pixels with each increment in resolution, compared to two new pixels in Section \ref{sec-wavelets-background}, is simply a consequence of the Haar wavelets being extended from the one-dimensional case to the two-dimensional case. The number of pixels at resolution $j$, denoted ${N_{j}}$, is therefore given by ${12 \times 4^{j - 1}}$. The Haar scaling function ${\phi_{j, k}(\bm{\hat{s}})}$, and wavelets ${\psi^m_{j, k}(\bm{\hat{s}})}$, are shown in Fig.~\ref{fig-haar-wavelets}.

\begin{figure}
        \centering
	\includegraphics[clip, width=90mm]{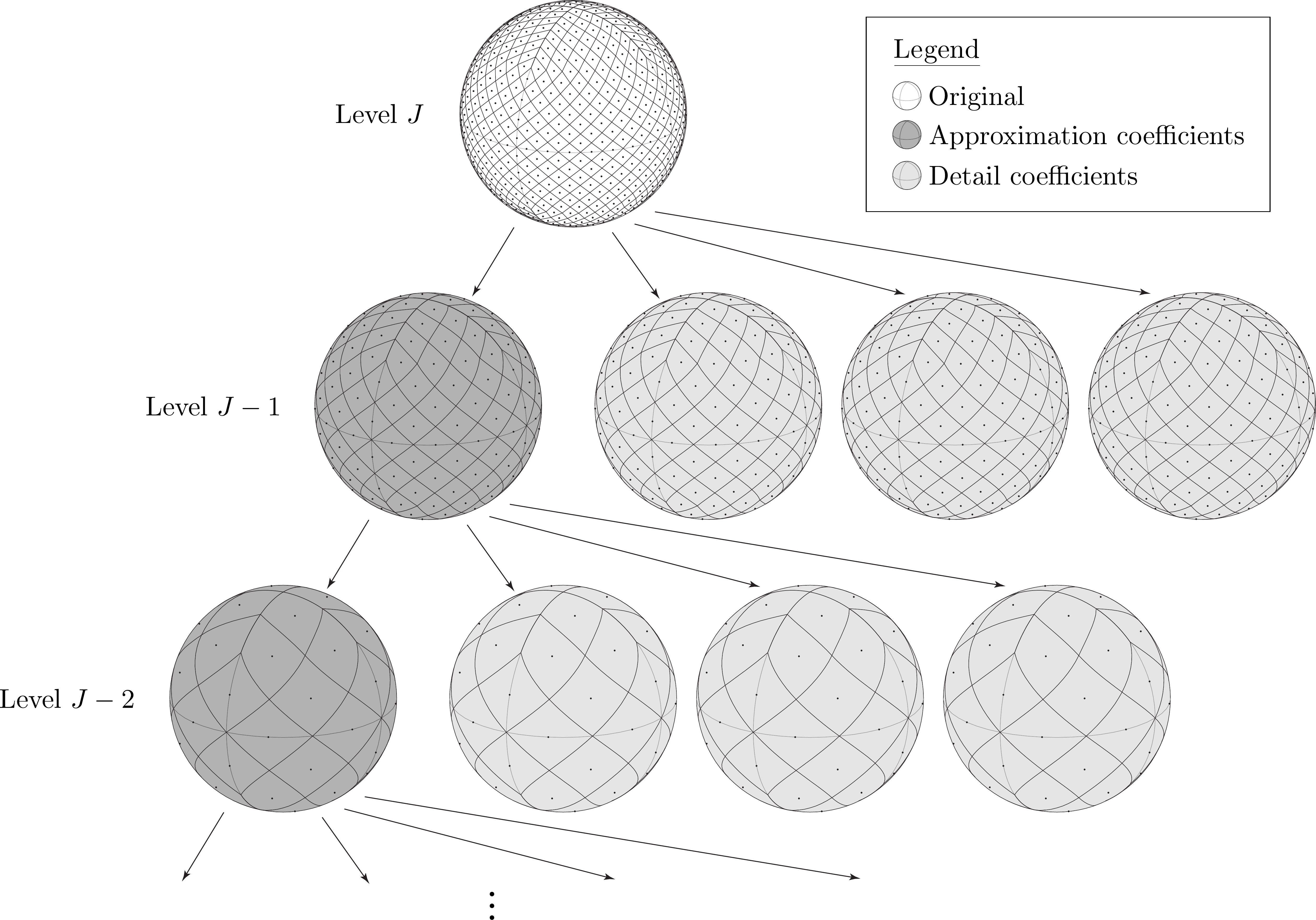}
        \caption{The recursive computation of scaling coefficients, ${\lambda_{j, k}}$, and wavelet coefficients, ${\gamma^m_{j, k}}$ (image from \citealt{McEwen2008}). The coefficients are first computed for resolution level $J$, and then recursively for levels ${j = J - 1}$, ${j = J - 2}$, etc. using Eqns.~\ref{eqn-scaling-coefficient} and \ref{eqn-wavelet-coefficient}.}
        \label{fig-haar-coefficients}
\end{figure}

To generate the scaling and wavelet coefficients we base our method upon the description in \cite{McEwen2008}, where the scaling coefficients, ${\lambda_{j, k}}$, for resolution $j$ and pixel $k$ are defined as
\begin{equation}
	\lambda_{j, k} = \frac{1}{4} \left( \lambda_{j + 1, 4k} + \lambda_{j + 1, 4k + 1} + \lambda_{j + 1, 4k + 2} + \lambda_{j + 1, 4k + 3} \right),
	\label{eqn-scaling-coefficient}
\end{equation}
and the wavelet coefficients, ${\gamma^{m}_{j, k}}$, of type $m$ are defined as
\begin{equation}
	\begin{split}
	\gamma^0_{j, k} &= \frac{\sqrt{A_j}}{4} \left( \lambda_{j + 1, 4k} - \lambda_{j + 1, 4k + 1} + \lambda_{j + 1, 4k + 2} - \lambda_{j + 1, 4k + 3} \right), \\
	\gamma^1_{j, k} &= \frac{\sqrt{A_j}}{4} \left( \lambda_{j + 1, 4k} + \lambda_{j + 1, 4k + 1} - \lambda_{j + 1, 4k + 2} - \lambda_{j + 1, 4k + 3} \right), \\
	\gamma^2_{j, k} &= \frac{\sqrt{A_j}}{4} \left( \lambda_{j + 1, 4k} - \lambda_{j + 1, 4k + 1} - \lambda_{j + 1, 4k + 2} + \lambda_{j + 1, 4k + 3} \right),
	\end{split}
	\label{eqn-wavelet-coefficient}
\end{equation}
where $A_j$ is the area of a single pixel at resolution $j$ in steradians. At the most detailed resolution, $j = J$, the scaling coefficients are populated directly from the value of the spherical function at that position (see Fig.~\ref{fig-haar-coefficients}), i.e. ${\lambda_{J, k} = F(\bm{\hat{s}}_k)}$, where ${\bm{\hat{s}}_k}$ is a unit vector pointing from the origin to the position of pixel $k$, and the coefficients for all levels ${j < J}$ are computed recursively using Eqns.~\ref{eqn-scaling-coefficient} and \ref{eqn-wavelet-coefficient}. Using these coefficients, full reconstruction of the original spherical function, ${F(\bm{\hat{s}})}$, is possible using (from \citealt{McEwen2008})
\begin{equation}
	F(\bm{\hat{s}}) = \sum^{N_1 - 1}_{k = 0} { \lambda_{1, k} \, \phi_{1, k}(\bm{\hat{s}}) } + \sum^{J - 1}_{j = 1} \sum^{N_j - 1}_{k = 0} \sum^2_{m = 0} { \gamma^m_{j, k} \, \psi^m_{j, k}(\bm{\hat{s}}) },
\end{equation}
where $N_j$ is the number of pixels at resolution level $j$, ${\phi_{j, k}(\bm{\hat{s}})}$ is the scaling function for pixel $k$ at resolution $j$, $J$ is the most detailed resolution used, and ${\psi^m_{j, k}}$ is the wavelet of type $m$ for pixel $k$ at resolution $j$. However, for radio imaging as presented here inversion is not required as the original spherical function, ${F(\bm{\hat{s}})}$, does not need to be recovered. Therefore, the scaling function, ${\phi_{j, k}}$, and wavelets, ${\psi^m_{j, k}}$, do not need to be computed explicitly.

\subsection{Spherical wavelet measurement equation}

During the major-cycle stage of Cotton-Schwab clean visibilities are normally recovered from the model image through the use of an FFT, followed by a de-gridding step, as described in Section \ref{sec-radio-imaging-cleaning}. In doing so, some degree of accuracy is sacrificed while interpolating over the uv grid, when compared to calculating visibilities directly using Eqn. \ref{eqn-visibility}. One could generate visibilities directly from the component list without the need to correct for the w-terms, and without sacrificing accuracy, using a variation of Eqn. \ref{eqn-visibility},
\begin{equation}
	V_{n} = \sum_{m = 0}^{N_{\rm C} - 1} {C_{m} \, {\rm exp}\big[ \, i2\pi \bm{B}_{n} \cdot \left( \bm{\hat{C}}_{m}^{\rm uvw} - \bm{\hat{P}} \right) \, \big]},
	\label{eqn-visibility-wavelet}
\end{equation}
where ${N_{\rm C}}$ is the number of distinct components identified during cleaning, ${B_n = [u_n, v_n, w_n]}$ is the baseline vector for visibility ${V_n}$, ${\bm{\hat{C}}_{m}^{\rm uvw}}$ is a unit vector pointing from the origin to the position of component ${C_m}$ on the sphere (where ${\bm{\hat{C}}_{m}^{uvw} = [c_m^{\rm u}, c_m^{\rm v}, c_m^{\rm w}]}$), and ${\bm{\hat{P}}}$ is a unit vector pointing from the origin to the phase position. The term 'distinct' is used in our definition of ${N_{\rm C}}$ as any clean components resolved to the same \textsc{healpix} pixel at the highest resolution are summed together in the spherical model image, so the run time scales with the number of pixels that have one or more clean components assigned to them.

In practice, the coordinate system used by ${\bm{\hat{C}}_m^{\rm uvw}}$ and ${\bm{\hat{P}}}$ will match that of ${\bm{B}_n}$ in orientation, and be aligned such that ${\bm{\hat{P}} = [0, 0, 1]}$, so the dot product in Eqn.~\ref{eqn-visibility-wavelet} becomes ${u_n c_m^{\rm u} + v_n c_m^{\rm v} + w_n \left( c_m^{\rm w} - 1 \right)}$.

A wavelet decomposition of Eqn. \ref{eqn-visibility-wavelet} is implemented based upon two spherical functions: the first represents the model image,
\begin{equation}
	F^{\rm I}(\bm{\hat{s}}) = \sum^{N_{\rm C} - 1}_{i = 0} C_i \, \delta( \bm{\hat{s}} - \bm{\hat{C}}^{\rm uvw}_i ),
\end{equation}
where $\delta$ is the Dirac delta function, and the second spherical function represents the plane wave,
\begin{equation}
	F^{\rm P}( \bm{\hat{s}}, \bm{B} ) = {\rm exp}\Big[ i2\pi \bm{B} \cdot {\left( \bm{\hat{s}} - \bm{\hat{P}} \right)} \Big],
\end{equation}
\begin{figure}
        \centering
	\includegraphics[clip, width=90mm]{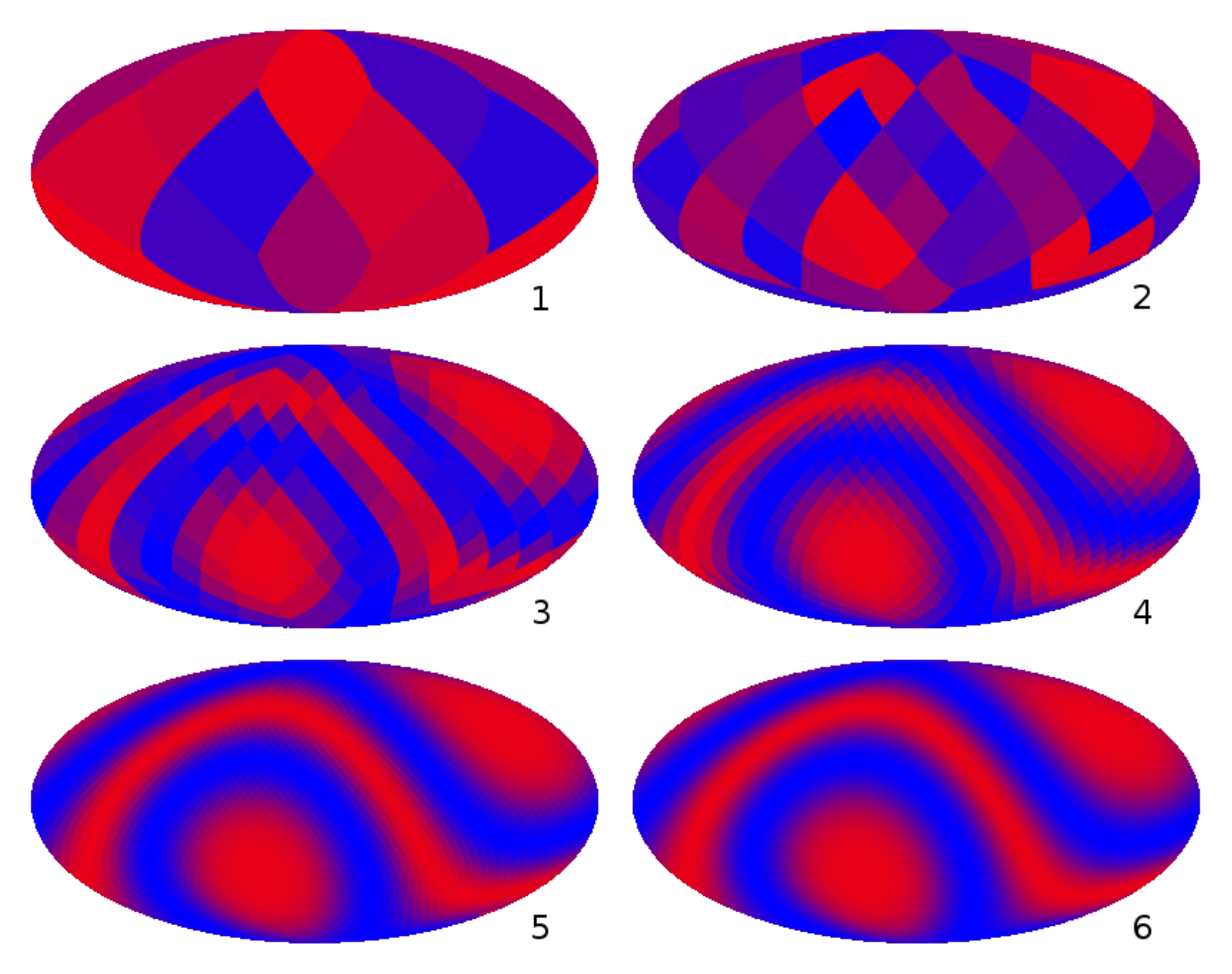}
        \caption{Scaling coefficients, ${\eta_{j, k}(\bm{B})}$, of the plane-wave function, ${F^{\rm P}(\bm{\hat{s}}, \bm{B})}$, for the first six resolution levels (${j = 1}$ to ${j = 6}$), shown over the whole unit two-sphere in Mollweide projection.}
        \label{fig-plane-wave}
\end{figure}
which is shown graphically in Fig.~\ref{fig-plane-wave} for the wavelet resolution levels ${j = 1}$ to ${j = 6}$. The visibility equation becomes an integration over the whole unit two-sphere $S^2$,
\begin{equation}
	V_i = \int_{S^2} { F^{\rm I}(\bm{\hat{s}}) \, F^{\rm P}( \bm{\hat{s}}, \bm{B}_{i} ) } \, {\rm d}\Omega(\bm{\hat{s}}),
\end{equation}
where ${{\rm d}\Omega(\bm{\hat{s}}) = {\rm sin}\theta \, {\rm d}\theta \, {\rm d}\phi}$. This integration can be expressed in wavelet form as a summation involving only the non-zero scaling and wavelet coefficients.

The spherical equivalent of the model image, ${F^{\rm I}(\bm{\hat{s}})}$, is constructed using a SIN projection of the clean components from the image plane to the surface of the sphere. For a clean component $C_i$ at image-plane position ${[ l_i, m_i ]}$ the position vector is given by ${\bm{\hat{C}}^{\rm uvw}_i = \Big[ { l_i, m_i, \sqrt{(1 - l_i^2 - m_i^2)} } \Big]}$. We hereafter use $\lambda_{j, k}$ and ${\gamma^{m}_{j, k}}$ to represent the scaling and wavelet coefficients of the model image, ${F^{\rm I}(\bm{\hat{s}})}$, and ${\eta_{j, k}(\bm{B})}$ and ${\sigma^m_{j, k}(\bm{B})}$ to represent the scaling and wavelet coefficients of the plane wave, ${F^{\rm P}(\bm{\hat{s}}, \bm{B})}$; \cite{McEwen2008} used ${\delta^m_{j, k}}$ to represent the plane-wave wavelet coefficients, but we choose to use $\sigma^m_{j, k}$ to distinguish this function from the Dirac delta function. The visibilities can be reconstructed directly from the resolution-$J$ scaling coefficients, using
\begin{equation}
	V_i = \sum^{N_J - 1}_{k = 0} {\lambda_{J, k} \, \eta_{J, k}\left( \bm{B}_i \right)}.
\end{equation}

It can be shown \citep[see][]{McEwen2008} that the orthogonality of the scaling and wavelet functions for spherical Haar wavelets leads to an expression for the visibilities in terms of the scaling and wavelet coefficients, given by
\begin{equation}
	V_i = \sum^{N_1 - 1}_{k = 0} { \lambda_{1, k} \, \eta_{1, k}( \bm{B}_i ) } + \sum^{J - 1}_{j = 1} \sum^{N_j - 1}_{k = 0} \sum^2_{m = 0} { \gamma^m_{j, k} \sigma^m_{j, k}( \bm{B}_i ) }.
\end{equation}

\begin{figure}
        \centering
	\includegraphics[clip, trim=1cm 2.8cm 5.8cm 8cm, width=110mm]{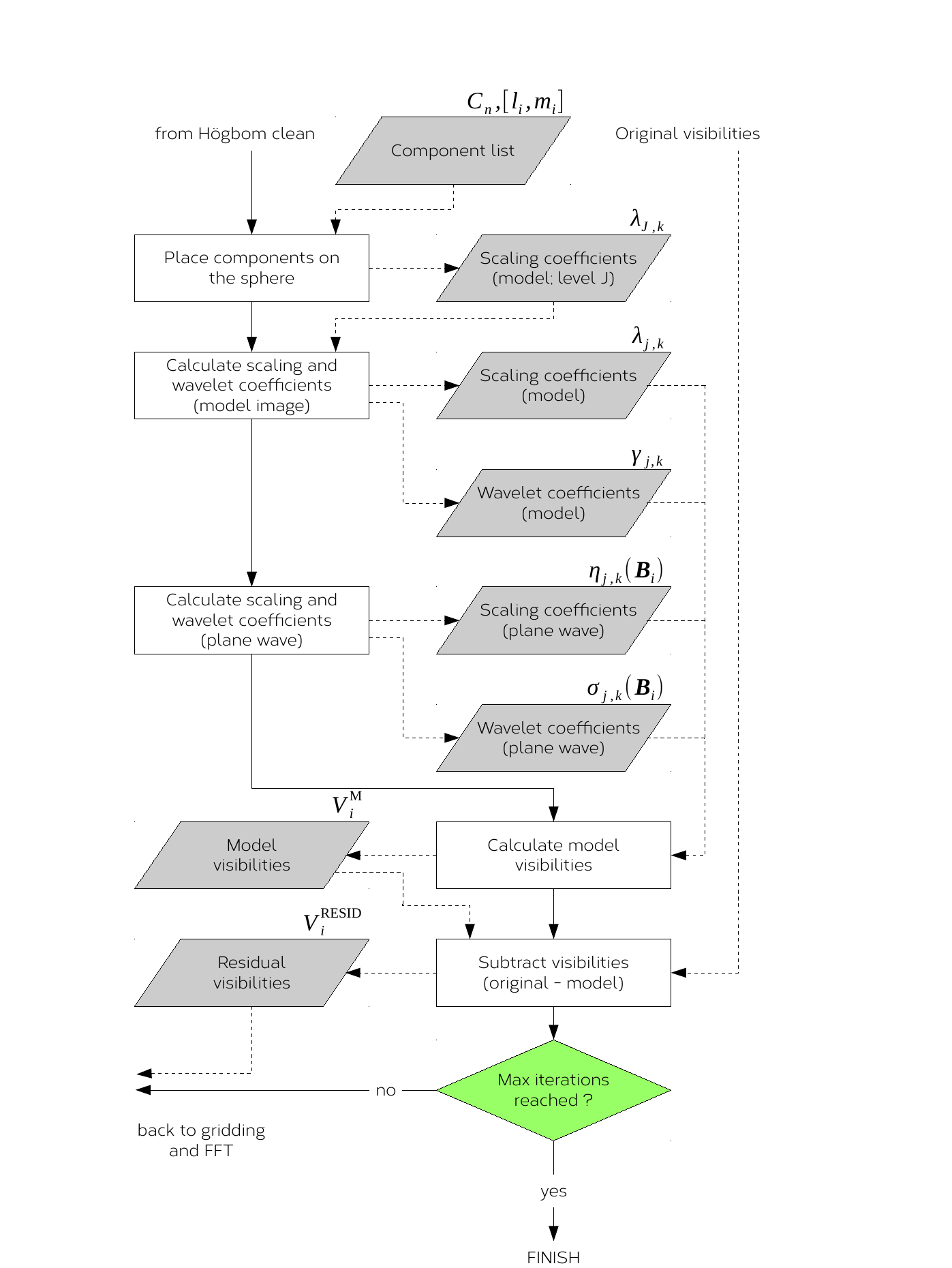}
        \caption{The major cycles of the wavelet clean transfer the clean components onto the sphere, and generate a set of scaling and wavelet coefficients from which the model visibilities can be calculated. The clean components directly set the model-image scaling coefficients at resolution $J$ (the most detailed resolution used), and the remainder of the model-image scaling and wavelet coefficients are calculated recursively for levels ${J - 1}$, ${J - 2}$, etc. A new set of plane-wave coefficients must be generated for each visibility.}
        \label{fig-flow-chart-wavelet}
\end{figure}

Starting with the component list generated by H\"ogbom clean, $C_i$ and ${[l_i, m_i]}$, the wavelet clean will construct the resolution-$J$ scaling coefficients for the model image and the plane wave (${\lambda_{J, k}}$ and ${\eta_{J, k}(\bm{B})}$ respectively), and recursively calculate the remaining coefficients ${\lambda_{j, k}}$, ${\gamma^m_{j, k}}$, ${\eta_{j, k}(\bm{B})}$ and ${\sigma^m_{j, k}(\bm{B})}$ for all resolutions $j$. Only one set of coefficients is calculated for the model image, but for the plane wave function, which is dependent upon the baseline vector, a separate set of coefficients must be generated for each visibility. A flow chart of the whole process is shown in Fig.~\ref{fig-flow-chart-wavelet}.

\begin{figure}
        \centering
	\includegraphics[clip, width=90mm]{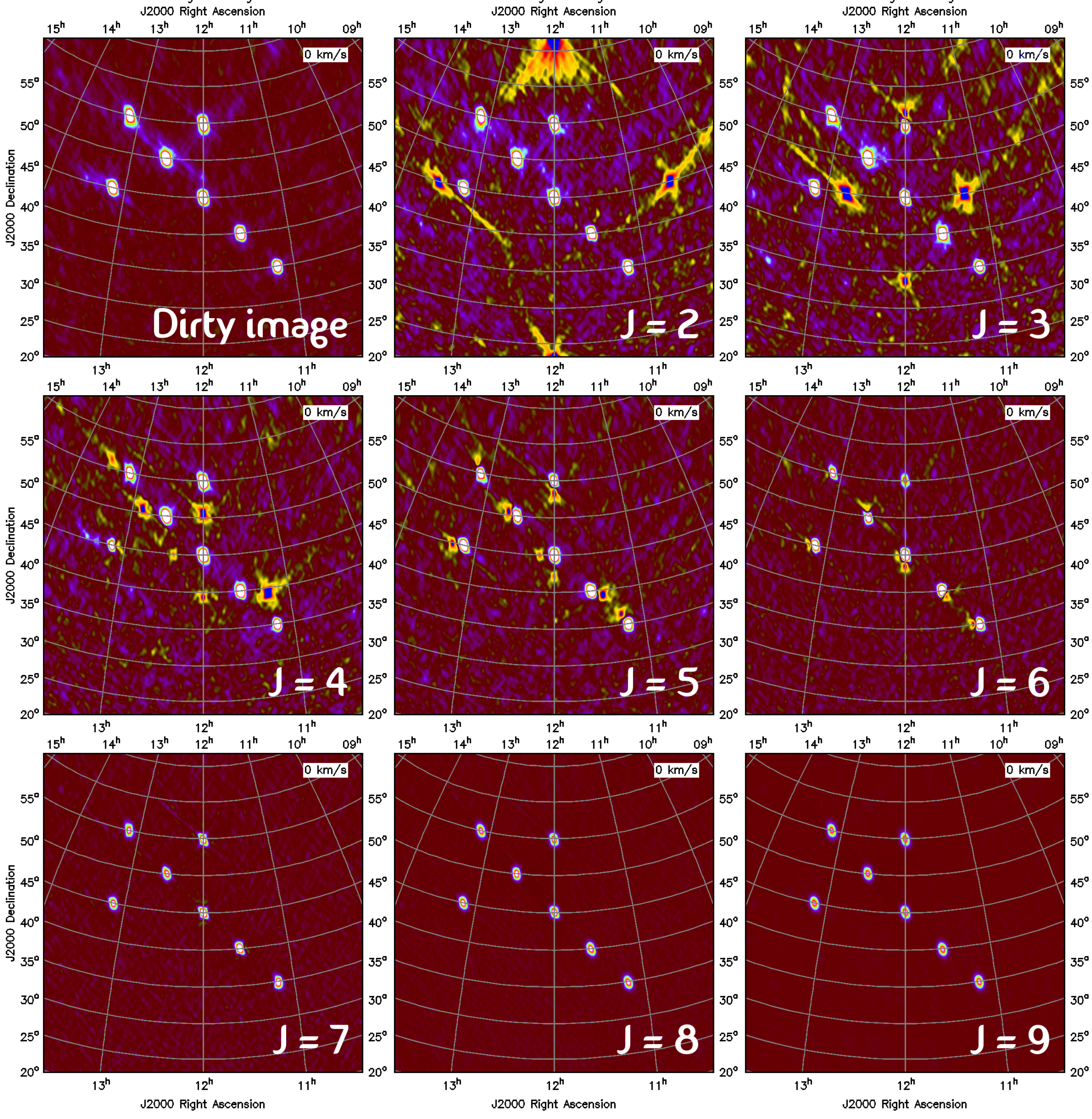}
        \caption{Simulated measurement set with an arrangement of seven compact sources centred at the phase position of ${\alpha = 12{\rm h}, \delta = 45^{\circ}}$, based upon radio antennae with Gaussian primary beams, arranged in the VLA-C configuration. The total observation time was 2\,016~s, and all sources were set to a flux density of 1~Jy. The observation frequency was 10~MHz (unphysically low for the VLA, but chosen for a large field of view and low angular resolution), and the field of view $42^{\circ}40'$. The dirty image (top-left panel) was cleaned with 600 CLEAN iterations (minor cycles) over two major cycles, and the remaining panels show the clean images when the resolution parameter was increased from $J = 1$ to $J = 9$.}
        \label{fig-simulated-data}
\end{figure}

It is essential when reconstructing visibilities from the model image that the resolution of our wavelet representation is sufficiently high that no significant loss of astrometric precision occurs. Fig.~\ref{fig-simulated-data} demonstrates the effect of setting the resolution parameter $J$ too low, using a simulated measurement set based upon the Karl Jansky Very Large Array (JVLA) in 'C' configuration. Whereas the clean image constructed using resolution ${J = 9}$ shows a deep contrast between the seven point sources and the background noise, the clean images constructed using lower values of $J$ show higher levels of background noise, and contain patches of negative flux density at the positions where the clean components have been placed in the spherical representation of the model image. These dark patches are seen to migrate towards the correct component positions as $J$ is increased.

The angular size of each pixel on the sphere (given by ${\Delta \theta_j \sim \sqrt{A_j} = \sqrt{4 \pi / N_j} = 2^{1 - j} \sqrt{\pi / 3}}$) should be no larger than the angular size of each pixel in the dirty image, and preferably the spherical representation of the model image should oversample the 2-D representation by a factor of two. Table~\ref{tbl-recommended-resolutions} shows the best angular resolution that can be achieved with a variety of well-known radio telescopes and configurations, and the recommended wavelet resolutions.

\begin{table}
	\footnotesize
	\caption{Recommended wavelet resolution, $J$, for a variety of well-known radio telescopes and configurations. The wavelet resolution was chosen so that each pixel on the sphere oversamples the pixels in the 2-D model image by a factor of at least two.}
	\label{tbl-recommended-resolutions}
	\begin{center}
	\begin{tabular}{@{}lcccccccc}
		\hline
		Telescope & Angular resolution & $J$ & Wavelet resolution \\
		& ($\Delta \theta$) & & (${\Delta \theta_J}$) \\
		& [milliarcsec] & & [milliarcsec] \\
		\hline
		JVLA, L-band, A-config & 1300$^a$ & 20 & 400 \\
		JVLA, X-band, A-config & 200$^a$ & 23 & 50 \\
		e-MERLIN, L-band & 150$^b$ & 23 & 50 \\
		e-MERLIN, K-band & 12$^b$ & 27 & 3.1 \\
		EVN, 18 cm (L-band) & 15$^c$ & 26 & 6.3 \\
		EVN, 3.6 cm (X-band) & 3$^c$ & 29 & 0.79 \\
		\hline
	\end{tabular} \\
	\end{center}
	{
		\raggedright
		\scriptsize
		\textsc{Notes}: \\
		$^a$ JVLA Observational Status Summary 2018B (https://science.nrao.edu/facilities/vla/docs/manuals/oss/performance/resolution), \\
		$^b$ Cycle-7 e-MERLIN observations (http://www.e-merlin.ac.uk/observe/call\_cycle7.html), \\
		$^c$ EVN user guide (http://www.evlbi.org/user\_guide/res.html) \\
	}
\end{table}

\subsection{The coordinate system}

The preferred coordinate system for radio interferometric data, which we call the uvw coordinate system, is aligned such that the $w$-axis points at the phase position, $\bm{\hat{P}}$, the $u$-axis points east when the phase position is at its zenith, and the $v$-axis completes the cross product. The other coordinate used by our software, which we call the xyz coordinate system, is aligned with the Earth's axis, such that the $y$-axis points to the north celestial pole (NCP; declination, ${\delta = +90^{\circ}}$), and the $x$- and $z$-axes define the plane of the equator, with the $x$-axis pointed at right ascension (RA) ${\alpha = 6~{\rm hours}}$ and the $z$-axis pointed at RA ${\alpha = 0~{\rm hours}}$. The xyz coordinate system is used as an intermediate step when calculating the RA and declination of a position on the sphere.

The Euler matrices, ${{\rm \bm{R}}_x}$, ${{\rm \bm{R}}_y}$ and ${{\rm \bm{R}}_z}$, are used to perform rotations on the sphere, and the transformation from the uvw to the xyz coordinate system is given by
\begin{equation}
	{\rm \bm{R}_{uvw \rightarrow xyz}}( \alpha_{\rm P}, \delta_{\rm P} ) = {\rm \bm{R}}_y( \alpha_{\rm P} ) \, {\rm \bm{R}}_x( -\delta_{\rm P} ),
\end{equation}
where ${\alpha_{\rm P}}$ and ${\delta_{\rm P}}$ are the RA and declination of the phase position.

\subsection{The algorithm}

The recommended wavelet resolutions provided in Table~\ref{tbl-recommended-resolutions} require that the software is able to distinguish very large numbers of pixels on the sphere. For example, in L-band, the recommended resolution, $J$, is 20 for the JVLA (in A-configuration), 23 for e-MERLIN, and 26 for the European Very-long Baseline Interferometer Network (EVN), corresponding to ${3.3 \times 10^{12}}$, ${2.1 \times 10^{14}}$, and ${5.4 \times 10^{16}}$ unique pixels respectively.

However, the number of clean components is usually no greater than a few thousand, and the number of unique image pixels amongst those components no larger than a few hundred. Therefore, the vast majority of pixels on the sphere will have zero values, and can be ignored when constructing the plane wave function and the scaling and wavelet coefficients.

\begin{figure}
        \centering
	\includegraphics[clip, trim=1cm 23.6cm 5.8cm 1.2cm, width=100mm]{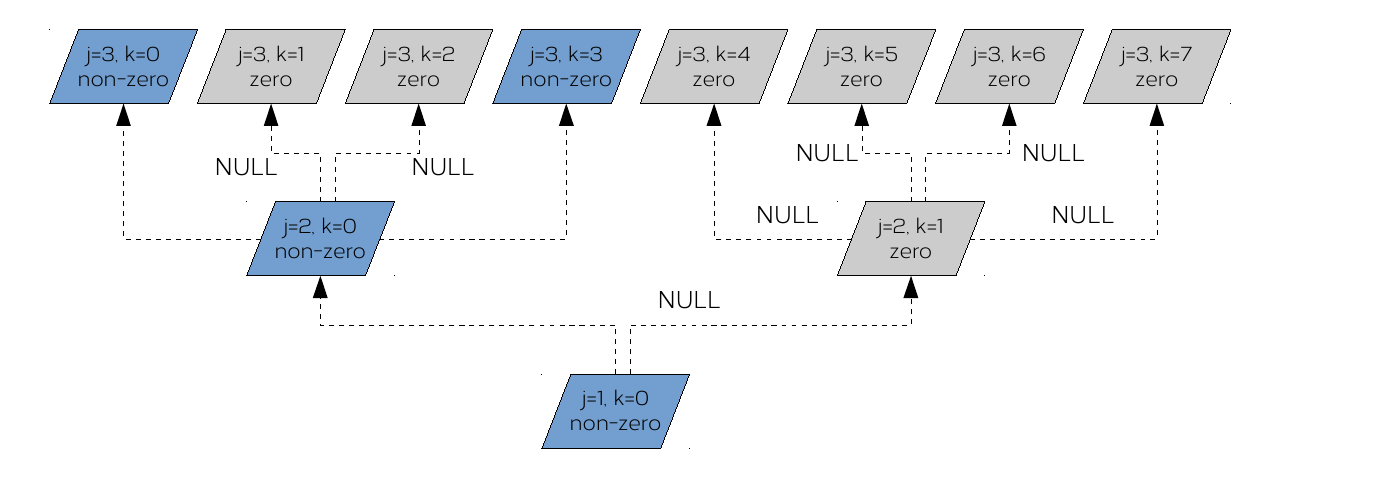}
        \caption{The scaling and wavelet coefficients are stored in a tree structure, with each node containing the coefficients for the model image and plane wave at resolution $j$, pixel $k$, as well as a pointer to up to four child nodes at resolution ${j + 1}$. Most of the pixels contain no clean components, so the plane wave and wavelet coefficients are not calculated for these branches.}
        \label{fig-flow-chart-tree}
\end{figure}

We represent our coefficients in a tree structure, as shown in Fig.~\ref{fig-flow-chart-tree}, where each node contains the scaling and wavelet coefficients of pixel $k$ at resolution $j$, and a pointer to up to four child nodes at resolution ${j + 1}$. A null pointer indicates that the child node, and all further nodes along this branch, are zero valued and can be ignored. The algorithms that work on these data, such as those that generate the plane wave and those that calculate the scaling and wavelet coefficients, work recursively, and starting from resolution ${j = 1}$ traverse the tree to find all non-zero nodes.

Before cleaning begins a cross-reference table is constructed that maps each image pixel ${[l, m]}$ to a pixel ${k_J}$ on the sphere at the highest resolution level $J$. From the image pixel ${[l, m]}$ we determine the position vector on the unit two-sphere, ${\Big[ { l, m, \sqrt{(1 - l^2 - m^2)} } \Big]}$, which is then transformed from the uvw coordinate system to the xyz coordinate system using the Euler rotation ${\rm \bm{R}_{uvw \rightarrow xyz}}$. It is then trivial to calculate the right ascension and declination, ${[\alpha_{l, m}, \delta_{l, m}]}$. The mapping is performed recursively, starting by identifying in which of the twelve pixels at level ${j = 1}$ we find the spherical coordinate ${[\alpha_{l, m}, \delta_{l, m}]}$, and then recursively moving down the tree by looking at the child nodes.

When a clean component, ${C_{i}}$, is identified at position ${[l_i, m_i]}$, the tree structure is traversed from ${j = 1}$ upwards. The node ID at level $J$ (denoted ${k_J}$) is known from the cross-reference table, and the node ID at all lower levels is given by ${k_j = k_J >> 2 (J - j)}$, where $>>$ denotes a bitwise shift to the right, i.e. an integer division by ${2^{2 (J - j)}}$. The nodes ${k_1}$ to ${k_J}$ are created, if they don't already exist, and the relevant pointers from parent to child nodes are set. When a new node is created at level $J$ the plane-wave function at this position on the sphere is calculated immediately. We do not calculate the scaling or wavelet coefficients at levels $j < J$ until all clean components have been found, and the full size of the tree is known.

\subsection{Performance}

\subsubsection{Run-time scaling}

The performance cost of the wavelet clean algorithm can be broken down into the steps of building the cross reference between the image pixels and the sphere pixels, generating the scaling and wavelet coefficients of the model image, generating the plane-wave function, generating the scaling and wavelet coefficients of the plane-wave function, and reconstructing the visibilities from the coefficients. The first two steps are performed only once per imaging run, and once per major cycle respectively, and therefore the major computational expense will be provided by the remaining steps, which must be performed once for every visibility.

Following H\"ogbom clean, the ${N_{\rm C}}$ distinct clean components, with values ${C_{i}}$ and image positions ${[l_i, m_i]}$, are cross referenced to positions on the sphere at resolution $J$, resulting in ${\widetilde{N}_{\rm C}}$ \textit{distinct} clean components. The process of traversing the tree to generate the plane wave, computing the scaling and wavelet coefficients, and reconstructing the visibilities all scale as ${\mathcal{O}(\widetilde{N}_{\rm C})}$, for each baseline.

\begin{figure}
        \centering
	\includegraphics[clip, width=90mm]{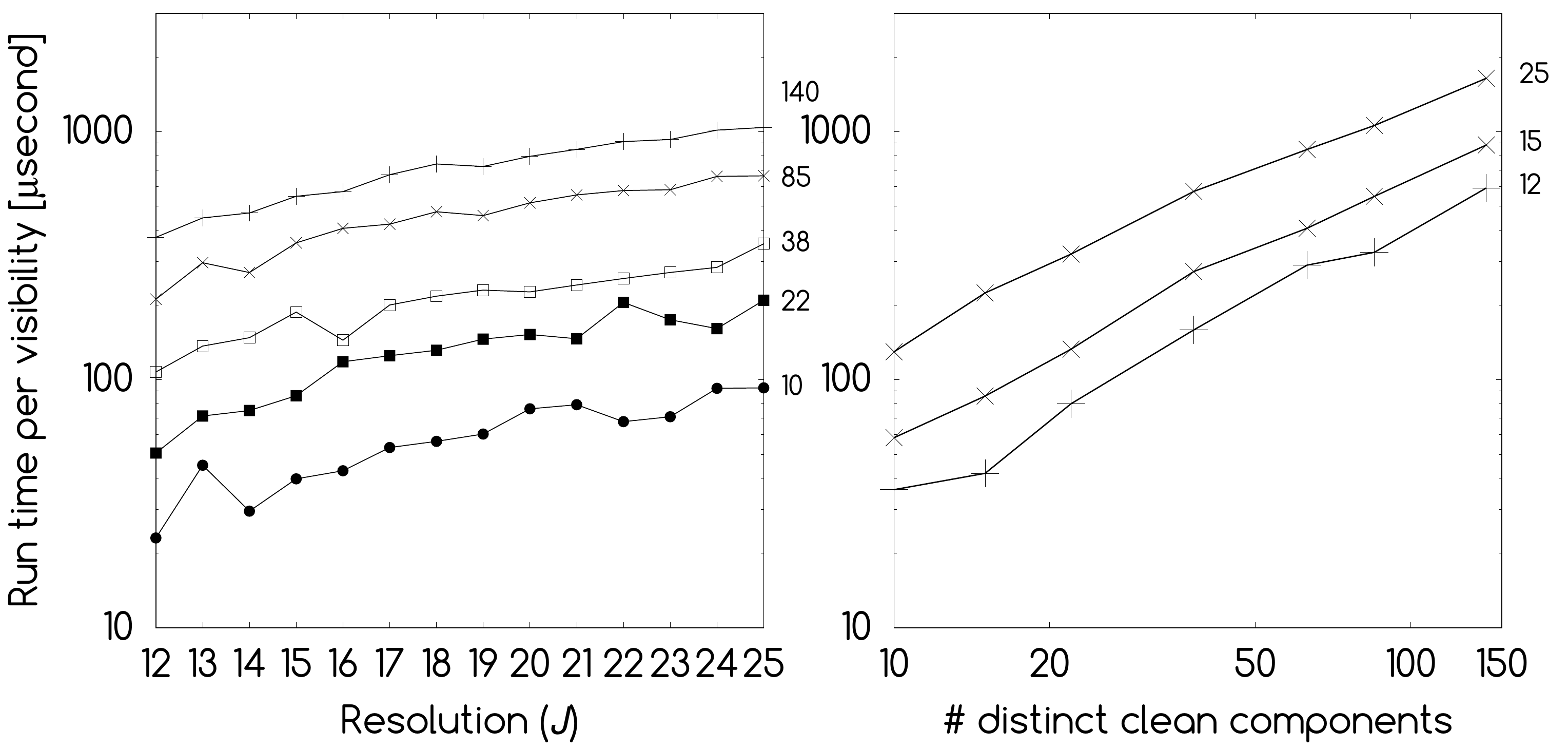}
        \caption{The execution time of wavelet clean, in microseconds per visibility. Left: the resolution, $J$, is varied between 12 and 25, and the performance is shown for fixed numbers of dinstinct clean components, ${\widetilde{N}_{\rm C}}$, which are indicated on the right-hand side of the panel. Right: the number of distinct clean components is varied between 10 and 140, and the performance is shown for ${J = 12}$, ${J = 15}$, and ${J = 25}$, also indicated to the right of the panel.}
        \label{fig-performance}
\end{figure}

In Fig. \ref{fig-performance} we show the time taken, per major cycle, by the wavelet clean prototype to process each visibility in the same simulated measurement set used in Fig. \ref{fig-simulated-data}. In the left panel, the resolution, $J$, was varied between 12 and 25, and the algorithm performance was found empirically to scale exponentially as ${\mathcal{O}(1.07^{J})}$, such that an increase in resolution from ${J = 12}$ to ${J = 25}$ results in only a ${\simnot2.4}$-times increase in execution time, and delivers a ${\simnot8\,200}$-times improvement in angular resolution. In the right panel, the number of minor cycles was adjusted such that the number of distinct clean components, ${\widetilde{N}_{\rm C}}$, varied between 10 and 140, and the performance was found to scale as ${\mathcal{O}(\widetilde{N}_{\rm C})}$.

Although generating the plane-wave function required using sine and cosine functions, this stage was responsible for only 5 per cent of the compute time demanded by each visibility. The dominant computational expense was the generation of the scaling and wavelet coefficients at resolution levels ${j = 1,...,J - 1}$, which took 60 per cent of processing time. The process of reconstructing the visibilities accounts for the remaining 35 per cent.

A one-hour observation using the JVLA, with a dump time of one second, generates a total of 1.26~million visibilities. A suitable resolution for an L-band, A-configuration observation would be ${J = 20}$, and generating wavelet coefficients, and reconstructing visibilities, from such a data set would take 1\,000~s, per major cycle, if 140 distinct clean components were found, or 146~s if only 15 distinct clean components were found.

\subsubsection{Image quality}

\begin{figure*}
        \centering
	\includegraphics[clip, width=140mm]{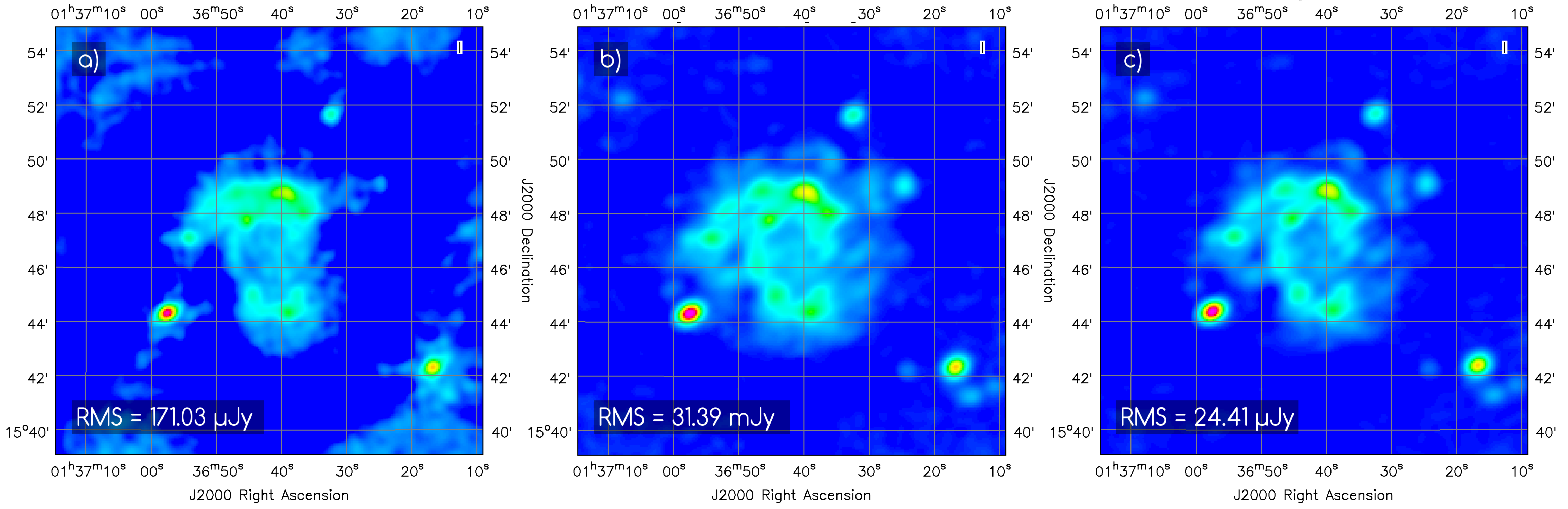}
        \caption{a) Dirty image constructed from a JVLA (D-config) observation of NGC~628 (from 13 March 2013, project 13A-014), generated using natural weighting. b) Image cleaned with 20\,000 CLEAN iterations using a conventional CLEAN algorithm that includes degridding. c) Image cleaned with 20\,000 iterations using wavelet clean.}
        \label{fig-normal-v-wavelet-clean-ngc628}
\end{figure*}

NGC~628 (M74) is a face-on spiral galaxy located at a distance of ${\rm\simnot7.3~Mpc}$ \citep{Sharina1996}. In Fig.~\ref{fig-normal-v-wavelet-clean-ngc628} we reduce a 20~kilosecond (ks), 2.5~GHz S-band JVLA observation (in D-configuration) of NGC~628, using a cell size of 3.7~arcsec, and a total of 3\,000 minor cycles (see \cite{Mulcahy2017} for a comprehensive discussion of this source and data set). The wavelet resolution was set to ${J = 18}$, giving 1.6~arcsec resolution on the sphere. The panels show the central 256 x 256 pixel regions of the dirty image (panel a), and clean images (using 20\,000 iterations; panel b is cleaned with a conventional CLEAN algorithm, including degridding in the major-cycle stage, and panel c is cleaned with wavelet clean), along with the noise rms as measured in a 400 x 400 pixel region of the images that is free from bright sources.

For these images Wavelet clean produced a lower rms (${\rm 24.41~\upmu Jy}$) compared to the algorithm using conventional cleaning (${\rm 31.39~\upmu Jy}$), representing a 22~per~cent reduction in noise. Since wavelet clean does not rely upon a flat-plane approximation of the sky when generating model visibilities it follows that the benefits to using this approach are expected to depend heavily upon the FOV of the image.

We have therefore constructed a simulated measurement set (see also Fig. \ref{fig-simulated-data}) based upon the JVLA in 'C' configuration. The data is a single channel and single polarisation observation of seven compact sources centred on the phase position ${\alpha = 12{\rm h}}, {\delta = 45^{\circ}}$. We have used an unphysically low observation frequency of 10~MHz, and no primary beam pattern, in order that a large FOV can be tested ($\simnot {40^{\circ}}$).

\begin{figure*}
        \centering
	\includegraphics[clip, width=180mm]{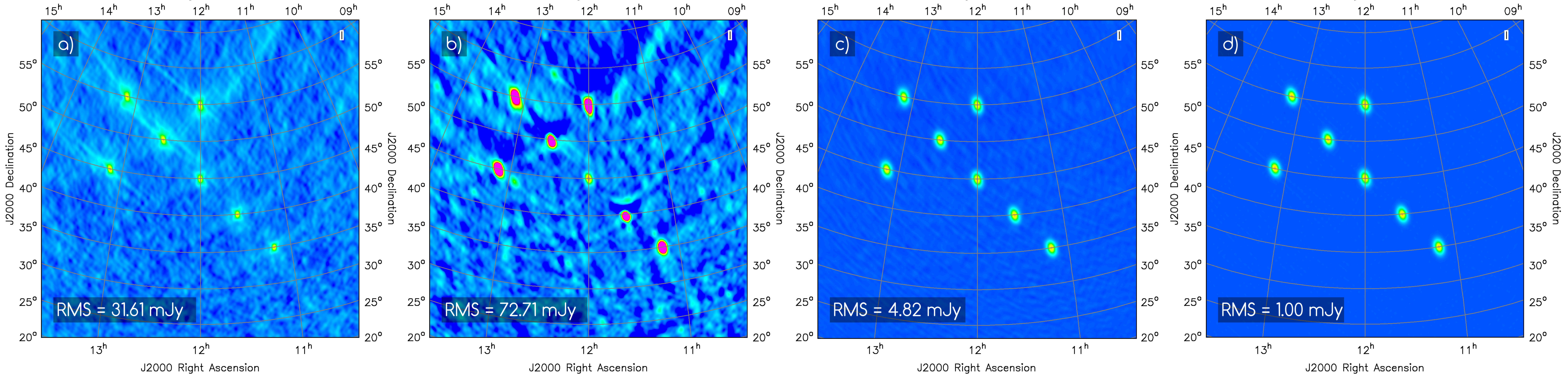}
        \caption{a) Dirty image from a wide-field simulated measurement set (See Fig. \ref{fig-simulated-data} for description), generated using 64 w-planes. b) Image cleaned with 700 CLEAN iterations over four major cycles using a conventional CLEAN algorithm that includes a forward FFT and degridding. For degridding we used only the anti-aliasing kernel, whereas both the anti-aliasing kernel and w-kernel (64 w-planes) were used during gridding. c) Image cleaned with 700 CLEAN iterations over four major cycles using a conventional degridding algorithm to calculate the model visibilities. We have used w-projection, with 64 w-planes, in both the forward and inverse directions. d) The same measurement set, cleaned with 580 CLEAN iterations over four major cycles, and using 64 w-planes while gridding and the wavelet clean algorithm in place of degridding.}
        \label{fig-normal-v-wavelet-clean}
\end{figure*}

Fig.~\ref{fig-normal-v-wavelet-clean} compares the dirty image (panel a), constructed over a 1024 x 1024 pixel grid using natural weighting and 64 w-planes, with the clean images produced by cleaning using a conventional CLEAN algorithm (panels b and c; the latter includes the w-kernel in the degridding stage, whereas the former has just the anti-aliasing kernel), and the clean image produced by wavelet clean (panel d, using ${J = 10}$). The root-mean-square (rms) noise, which is computed for each image from a region of 400 by 400 pixels at the bottom-left corner, was found to be 1.00~mJy for the wavelet-cleaned image (at ${J = 10}$), a factor of nearly 5x improvement over that of the image obtained using degridding (4.82~mJy).

\subsection{Comparison with CASA}

The Common Astronomy Software Application (CASA) \footnote{https://casa.nrao.edu/} \citep{McMullin2007} is a well-established tool for generating cleaned images from radio interferometric data. Combining gridding, fft, and cleaning into a single task, CASA is a benchmark against which the performance of wavelet clean can be assessed.

We hereafter use the term 'CASA clean' to represent the 'clean' method of the imager tool in CASA. We first use the 'defineimage' method to set the image size and cell size, and to use the multifrequency synthesis (mfs) mode. If w-projection is used then the parameters are set using the 'setoptions' method. Finally, cleaning is performed using the 'clean' method with the Cotton-Schwab algorithm, and a loop gain of 0.1.

\begin{figure}
        \centering
	\includegraphics[align=t,clip, width=60mm]{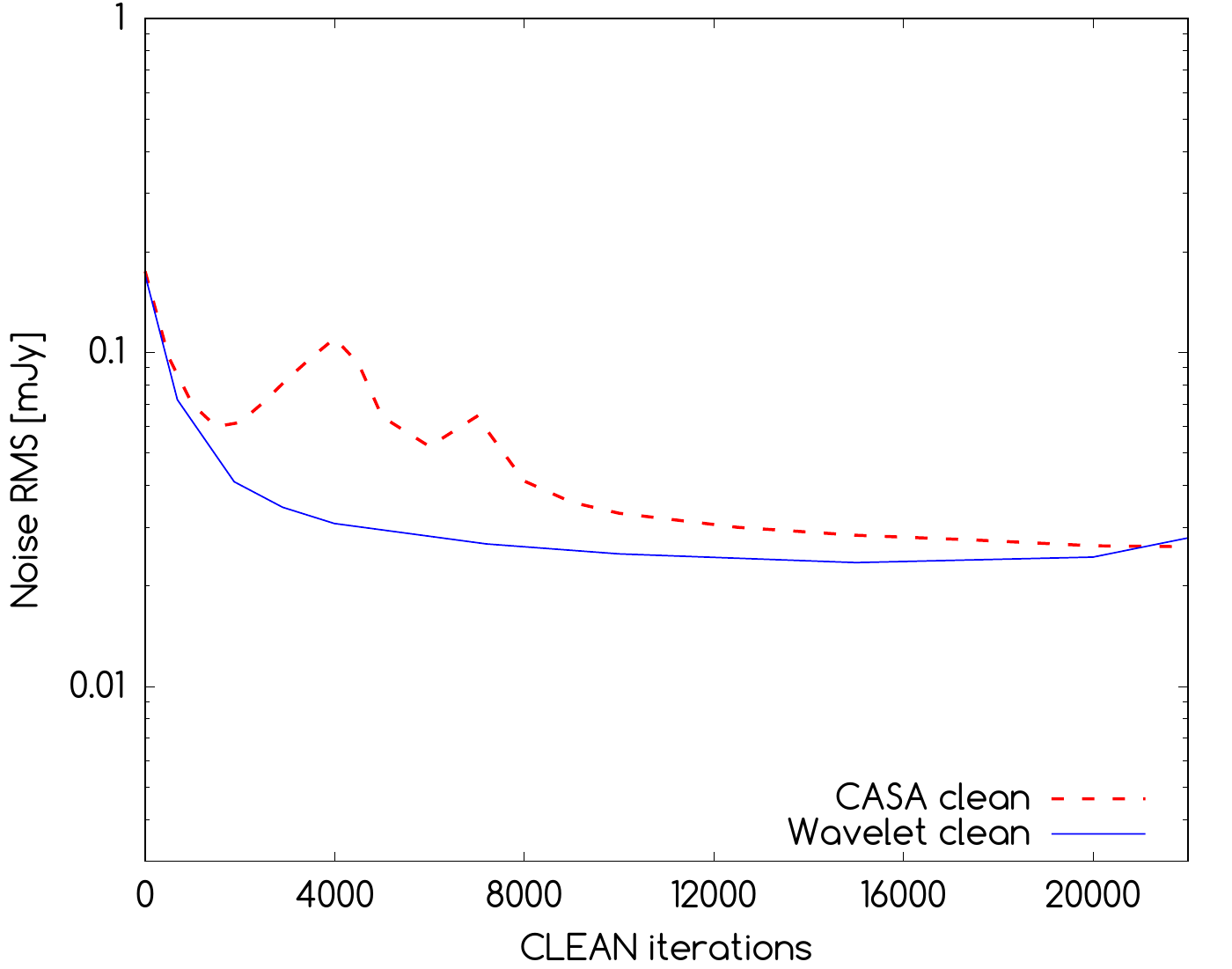}
	\includegraphics[align=t,clip, width=27mm]{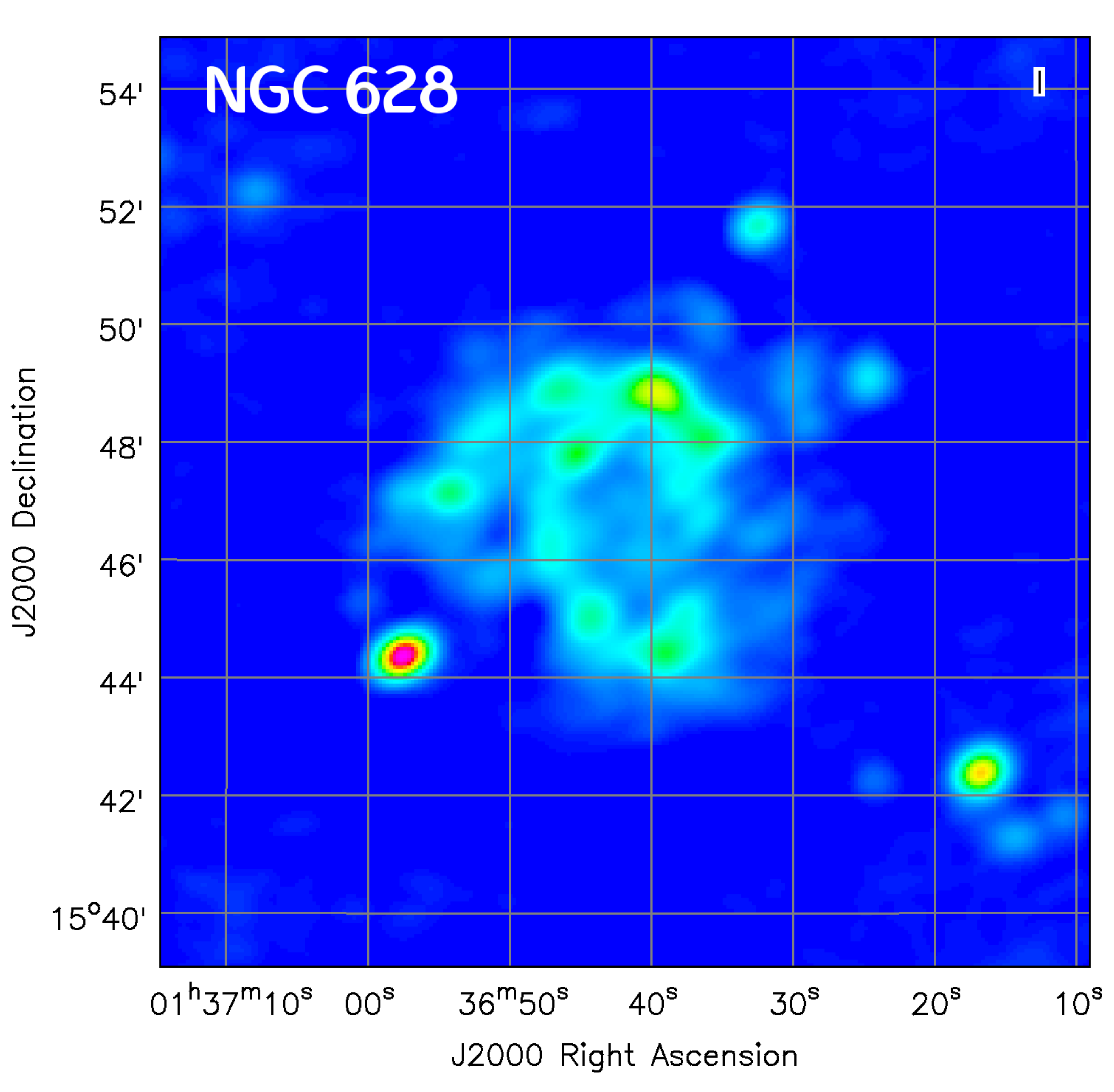}
	\includegraphics[align=t,clip, width=60mm]{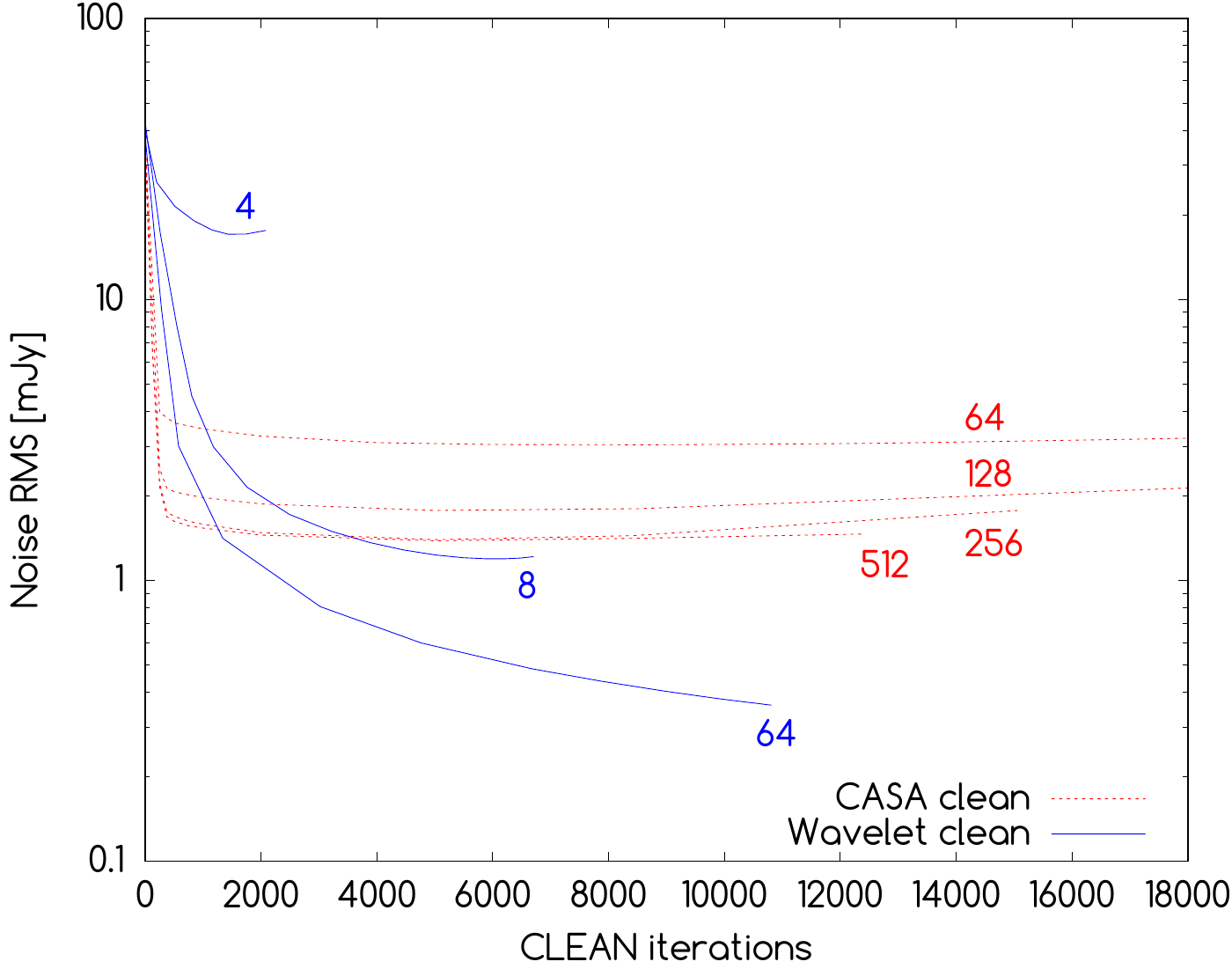}
	\includegraphics[align=t,clip, width=27mm]{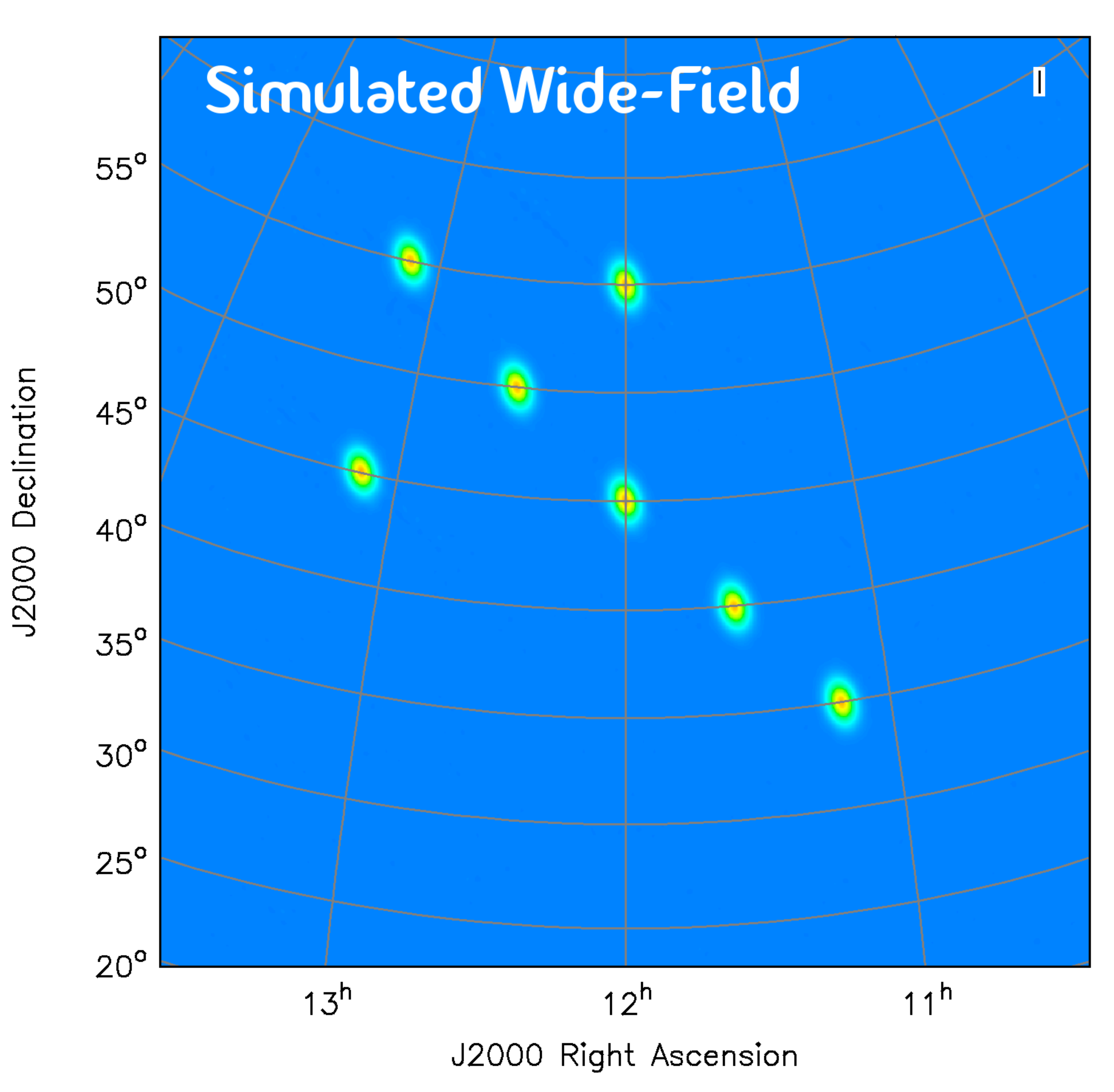}
        \caption{Top: The rms of wavelet clean and CASA clean images, based upon the data for NGC~628, as a function of the number of CLEAN iterations. We used natural weighting, a cell size of 3.7~arcsec, and a wavelet resolution of ${J = 18}$. The dashed red lines represent CASA clean, and the solid blue lines wavelet clean. Bottom: The rms of wavelet clean and CASA clean images, based upon the same simulated wide-field data set shown in Fig. \ref{fig-normal-v-wavelet-clean}. We used natural weighting, a cell size of 150~arcsec, and a wavelet resolution of ${J = 10}$. The number of w-planes used is indicated on the plot (for wavelet clean, w-projection was used in gridding only).}
        \label{fig-casa-v-waveletclean}
\end{figure}

In Fig.~\ref{fig-casa-v-waveletclean} we plot the rms noise of images from two datasets, each cleaned using CASA clean and wavelet clean, as a function of the number of minor cycles. The top panel shows the rms of the images of NGC~628 (see Fig.~\ref{fig-normal-v-wavelet-clean-ngc628}), and the bottom panel shows the rms of the simulated wide-field data set from Fig.~\ref{fig-normal-v-wavelet-clean}.

\subsubsection{NGC~628}

For NGC~628 we used natural weighting and a wavelet resolution of ${J = 18}$ (1.6~arcsec angular resolution). The image sizes are 1024 by 1024 pixels in size, and the cell size is 3.7~arcsec per pixel. The rms was calculated from a 400 by 400 pixel region of the clean image in which no bright sources are found. The wavelet clean algorithm and CASA clean tasks were found to clean to a comparable level of noise (the rms is ${\rm 23.5~\upmu Jy}$ and ${\rm 27.2~\upmu Jy}$ respectively). We do not at this stage compare the respective run times of wavelet clean and CASA clean, except to note that the execution time of our current implementation of wavelet clean is considerably longer than that of CASA clean.

\subsubsection{Simulated wide-field data}

For the simulated wide-field data we constructed clean images that are 1024 by 1024 pixels in size, have angular sizes of 42~deg, 40~minutes along each axis, and a 150~arcsecond cell size. W-projection was used for both CASA clean and wavelet clean (gridding only), and the number of w-planes used is indicated on the plot next to each curve.

As the number of w-planes is increased the CASA cleaned images show a diminishing return in the improvement of the rms noise, ultimately levelling out at a similar noise level to that found in the wavelet cleaned images that used eight w-planes during gridding. When 64 w-planes were used for both wavelet clean and CASA clean we find the rms noise of wavelet clean images to be a factor of more than nine times lower than those of CASA clean images (${\rm 359~\upmu Jy}$ and ${\rm 3.29~mJy}$ respectively.)

\section{Conclusions}

Generating model visibilities directly from a component list is a more accurate, but potentially computationally expensive, means of cleaning radio images, when compared to the more traditional method of performing a forward FFT and degridding. Forthcoming radio interferometers, such as the Square Kilometer Array (SKA), will not only offer large fields of view, and require large images (thousands of megapixels), but also generate extremely large numbers of visibilities (a few million per second). Although generating model visibilities directly eliminates the need for a forward FFT during each major cycle, we instead must sum over all items in the component list, for each visibility. Consequently, this method may be more efficient in cases where the number of clean components is small, and the size of the image is large. In this paper we have described our prototype which builds a full-sky model image from the component list, and decomposes this image into spherical wavelet coefficients.

The sparsity of the model images generated during cleaning makes a wavelet method of simulating visibilities a feasible option, whereas computing those visibilities using the spherical harmonic method is slower, and scales poorly \citep{McEwen2008}. Our prototype exploits this sparsity by only storing the non-zero coefficients, and navigating these components efficiently using a tree structure that links parent nodes directly to their child nodes.

We find that the image fidelity of wavelet-clean images is better than that of images cleaned with CASA, and the improvement in quality is particularly noticeable when the image FOV is large. For our wide-field (${\simnot 42^{\circ}}$ FOV) simulated dataset we observe a 8x improvement in the rms noise of the clean image (using the same number of w-planes whilst gridding), and a much smaller improvement of ${\rm \simnot 9.5~per~cent}$ in an S-band JVLA observation of NGC~628 (without w-projection).

\section*{Acknowledgements}

We thank David Mulcahy for providing the calibrated and flagged JVLA data on NGC~628.

\section*{References}

\bibliographystyle{model2-names}
\bibliography{references}

%% Authors are advised to submit their bibtex database files. They are
%% requested to list a bibtex style file in the manuscript if they do
%% not want to use model2-names.bst.

%% References without bibTeX database:

% \begin{thebibliography}{00}

%% \bibitem must have one of the following forms:
%%   \bibitem[Jones et al.(1990)]{key}...
%%   \bibitem[Jones et al.(1990)Jones, Baker, and Williams]{key}...
%%   \bibitem[Jones et al., 1990]{key}...
%%   \bibitem[\protect\citeauthoryear{Jones, Baker, and Williams}{Jones
%%       et al.}{1990}]{key}...
%%   \bibitem[\protect\citeauthoryear{Jones et al.}{1990}]{key}...
%%   \bibitem[\protect\astroncite{Jones et al.}{1990}]{key}...
%%   \bibitem[\protect\citename{Jones et al., }1990]{key}...
%%   \harvarditem[Jones et al.]{Jones, Baker, and Williams}{1990}{key}...
%%

% \bibitem[ ()]{}

% \end{thebibliography}

\end{document}